\documentclass[11pt]{article}

\usepackage[a4paper,margin=29mm]{geometry}
\usepackage[T1]{fontenc}
\usepackage{lmodern}
\usepackage{microtype}
\usepackage{amsmath,amssymb,amsthm,mathtools}
\usepackage{enumitem}
\usepackage{xcolor}
\usepackage{listings}
\usepackage[hidelinks]{hyperref}
\usepackage{listings}
\usepackage{xcolor}
\usepackage{amssymb}
\usepackage{pgfplots}
\usepackage[bottom]{footmisc}
\pgfplotsset{compat=1.18}

\definecolor{keywordcolor}{HTML}{0000FF}   
\definecolor{tacticcolor}{HTML}{0000FF}    
\definecolor{commentcolor}{HTML}{008000}   
\definecolor{stringcolor}{HTML}{A31515}    
\definecolor{backcolor}{HTML}{F8F8F8}      

\lstdefinelanguage{lean}{
  morekeywords=[1]{def, theorem, lemma, example, inductive, class, instance, structure, variable, variables, section, end, namespace, open, import, axiom, noncomputable, theory, with, without},
  keywordstyle=[1]\color{keywordcolor}\bfseries,
  morekeywords=[2]{by, exact, apply, intro, intros, simp, simp_all, rw, rewrite, calc, have, let, show, funext, unfold, congr, ext, ring, ring_nf, linarith, field_simp, filter_upwards, change, apply_rules},
  keywordstyle=[2]\color{tacticcolor},
  sensitive=true,
  morecomment=[l]{--},
  morecomment=[n]{/-}{-/},
  commentstyle=\color{commentcolor}\itshape,
  morestring=[b]",
  stringstyle=\color{stringcolor},
  columns=flexible,
  literate=
    {↦}{{$\mapsto$}}1
    {→}{{$\to$}}1
    {∀}{{$\forall$}}1
    {∃}{{$\exists$}}1
    {ℕ}{{$\mathbb{N}$}}1
    {ℝ}{{$\mathbb{R}$}}1
    {ℂ}{{$\mathbb{C}$}}1
    {∫}{{$\int$}}1
    {∂}{{$\partial$}}1
    {π}{{$\pi$}}1
    {γ}{{$\gamma$}}1
    {μ}{{$\mu$}}1
    {ε}{{$\varepsilon$}}1
    {ω}{{$\omega$}}1
    {≤}{{$\le$}}1
    {∈}{{$\in$}}1
    {≠}{{$\neq$}}1
    {∑}{{$\sum$}}1
    {•}{{$\bullet$}}1
    {ᵐ}{{$^m$}}1
    {��}{{$\mathcal{N}$}}1  
    {‖}{{$\|$}}1
    {₁}{{$_1$}}1
    {₂}{{$_2$}}1
    {↑}{{$\uparrow$}}1
    {∘}{{$\circ$}}1,
  basicstyle=\ttfamily\footnotesize,
  breaklines=true,
  keepspaces=true, 
  frame=single,
  rulecolor=\color{gray!30},
  backgroundcolor=\color{backcolor},
  numbers=left,
  numberstyle=\tiny\color{gray},
  stepnumber=1,
  tabsize=2,
  showstringspaces=false
}
\usepackage{graphicx}%
\usepackage{multirow}%
\usepackage{amsmath,amssymb,amsfonts}%
\usepackage{amsthm}%
\usepackage{mathrsfs}%
\usepackage[title]{appendix}%
\usepackage{xcolor}%
\usepackage{textcomp}%
\usepackage{manyfoot}%
\usepackage{booktabs}%
\usepackage{algorithm}%
\usepackage{algorithmicx}%
\usepackage{algpseudocode}%
\usepackage{listings}%

\usepackage{tikz}
\usetikzlibrary{decorations.markings}

\newtheorem{theorem}{Theorem}
\title{From the Dirichlet Integral to Lobachevsky’s Formula:
\\
A Formalization in Lean 4}

\author{
Daniel Goldberg\thanks{
  Department of Mathematics,
  Technion -- Israel Institute of Technology, Haifa, Israel.
  Email: \href{mailto:daniel.gold@campus.technion.ac.il}
  {\texttt{daniel.gold@campus.technion.ac.il}}
}
\and
Antoine Vinciguerra\thanks{
  Department of Computer Science,
  Technion -- Israel Institute of Technology, Haifa, Israel.
  Email: \href{mailto:antoine.v@campus.technion.ac.il}
  {\texttt{antoine.v@campus.technion.ac.il}}
}
}
\date{}
\begin{document}

\maketitle


\abstract{
We formalize the Dirichlet integral and several of its classical
applications in the Lean~4 proof assistant. Since the sinc function is
not Lebesgue integrable on the positive half-line, the Dirichlet
integral must be represented as the limit of integrals over bounded
intervals. To avoid the difficulty of removing an exponential
factor from a conditionally convergent integral, we instead pass
through the absolutely integrable function
\(\operatorname{sinc}^2\). We evaluate its integral by differentiation
under the integral sign and dominated convergence, and then recover the
Dirichlet integral from an identity between truncated integrals. Using
these results, we formalize the convergence of the Dirichlet cutoff to
the Heaviside function and derive several quadratic and bilinear
trigonometric integral identities. Finally, we formalize Lobachevsky's
integral formula for continuous periodic functions satisfying a
reflection symmetry, using the density of cosine polynomials obtained
from Mathlib's Fourier analysis on the additive circle.
}

\medskip 


\section{Introduction}\label{sec:introduction}

The Dirichlet integral is a classical example of a conditionally
convergent improper integral. Although its statement involves only
elementary functions, its evaluation already exhibits several ideas
that are central to real and Fourier analysis.

\subsection{The Dirichlet integral and its historical origins}

The Dirichlet integral is the identity
\begin{equation}\label{eq:dirichlet-integral}
    \lim_{T\to+\infty}
    \int_0^T \frac{\sin t}{t}\,dt
    =
    \frac{\pi}{2}.
\end{equation}

Throughout the paper, we use the continuous extension
\[
    \operatorname{sinc}(t)=
    \begin{cases}
        \sin(t)/t,&t\neq0,\\
        1,&t=0.
    \end{cases}
\]

The integral arose from the development of Fourier analysis. Fourier's
study of the heat equation introduced representations of functions by
trigonometric series \cite{fourier1822}, but their convergence was not
yet rigorously understood. In his 1829
memoir, Dirichlet gave the first general convergence theorem for
Fourier series under suitable regularity assumptions
\cite{dirichlet1829}.

By expressing the partial sums through what is now called the
Dirichlet kernel, Dirichlet reduced the convergence problem to the
study of oscillatory integrals. One of the central limits in his
argument has the form
\[
    \lim_{N\to+\infty}
    \int_0^h
        f(\beta)\frac{\sin(N\beta)}{\sin\beta}\,d\beta
    =
    \frac{\pi}{2}f(0),
\]
under appropriate continuity and monotonicity assumptions on \(f\)
near the origin. For \(f=1\), a rescaling of the integration variable leads to the Dirichlet integral.

The Dirichlet integral subsequently became a standard example of a
conditionally convergent improper integral
\cite{hardy1909integral,titchmarsh1948fourier}.

A concrete consequence is obtained from the normalized primitive
\begin{equation}\label{eq:dirichlet-primitive}
    S(x)
    :=
    \frac{1}{2}
    +
    \frac{1}{\pi}
    \int_0^x \operatorname{sinc}(t)\,dt.
\end{equation}
Equation~\eqref{eq:dirichlet-integral} and the evenness of
\(\operatorname{sinc}\) imply that
\[
    \lim_{R\to+\infty} S(Rx)
    =
    \begin{cases}
        0, & x<0,\\
        \dfrac{1}{2}, & x=0,\\
        1, & x>0.
    \end{cases}
\]
Thus, the rescaled primitive converges pointwise to the Heaviside
function. This illustrates the role of the Dirichlet integral in
Fourier inversion and in the approximation of discontinuous functions
\cite{zygmund2003trigonometric}.

The sinc function later became fundamental in harmonic analysis and
signal processing. Up to normalization, it is the inverse Fourier
transform of the indicator function of an interval and appears as the
interpolation kernel in the Shannon sampling theorem
\cite{shannon1949}.

Finally, the Dirichlet integral leads to further identities involving
sine and sinc functions, as well as to Lobachevsky's integral formula.
We formalize these results in the Lean~4 proof assistant
\cite{demoura2021lean}, using the Mathlib library
\cite{mathlib2020}. The complete formal development is available in
the accompanying repository \cite{dirichletgithub}.

\subsection{Challenges in the formalization}
The main difficulty in formalizing the Dirichlet integral lies not in
the evaluation of its value, but in the choice of an appropriate notion
of integral. In informal mathematics, the notation
\[
    \int_0^{+\infty}\operatorname{sinc}(t)\,dt
\]
is commonly used without further qualification. Since
\(\operatorname{sinc}\) is not absolutely integrable, however, this
expression cannot be interpreted as an ordinary Lebesgue integral. In
Mathlib, the Bochner integral of a non-integrable function is defined
to be zero. Therefore, the formal statement of the Dirichlet integral
must concern the convergence of the truncated integrals:
\[
    \int_0^T \operatorname{sinc}(t)\,dt
    \longrightarrow \frac{\pi}{2}
    \qquad\text{as }T\to+\infty.
\]
In Lean, this convergence is expressed using the filters
\texttt{atTop} and \texttt{Tendsto}. This formulation faithfully
represents conditional convergence, but it also prevents several
standard results about Lebesgue integrals from being applied directly.

This distinction also affects the classical argument based on the effect of the
exponential decay on the kernel. Although the integrals are absolutely
convergent, removing the exponential cannot be justified by a direct
application of the dominated convergence theorem: the natural pointwise majorant is
\(\lvert\operatorname{sinc}\rvert\), which is not integrable.

Our formalization therefore follows a different proof, in which the
conditionally convergent integral of \(\operatorname{sinc}\) is
recovered from the absolutely convergent integral of
\(\operatorname{sinc}^2\).

\subsection{Main contributions}

This paper presents a Lean~4 formalization of three closely related
families of results in real analysis:

\begin{itemize}
    \item \textbf{The Dirichlet integral.}
    We formalize the Dirichlet integral as the convergence of integrals
    over bounded intervals. 

    \item \textbf{Further integral identities.}
    We develop several consequences of the Dirichlet integral. In
    particular, we formalize the convergence of the Dirichlet cutoff to
    the Heaviside function, together with a collection of real and
    complex trigonometric integral identities.

    \item \textbf{Lobachevsky's integral formula.}
    We formalize Lobachevsky's formula for continuous periodic
    functions satisfying the appropriate symmetry assumptions. 
\end{itemize}
\subsection{Related formalization work and structure of the paper}
The Dirichlet integral was previously formalized in Isabelle/HOL by
Avigad, Hölzl, and Serafin as part of their formalization of the
central limit theorem \cite{avigad2017clt}. Their proof represents
\(1/x\) as an integral of \(e^{-ux}\), exchanges the resulting
integrals using Fubini's theorem, and applies dominated convergence
after evaluating the inner integral. Our formalization follows a
different route: it first evaluates the absolutely convergent integral
of \(\operatorname{sinc}^2\), and then recovers the Dirichlet integral
from an identity between truncated integrals. In both developments,
conditional convergence is expressed explicitly as a limit of
integrals over bounded intervals.

To the best of our knowledge, Lobachevsky's integral formula has not
previously been formalized in a proof assistant. The corresponding
part of our development therefore goes beyond a new formalization of
the Dirichlet integral: it provides a machine-checked proof of a
general identity for continuous periodic functions, combining the
integral identities established earlier with uniform approximation by
cosine polynomials.

The approximation argument builds on Mathlib's Fourier analysis on the
additive circle. This infrastructure has also been used in the
formalization of zeta and \(L\)-functions by Loeffler and Stoll
\cite{loefflerstoll2025}; our contribution here is its application to
Lobachevsky's formula rather than the underlying Fourier theory.

\paragraph{Structure of the paper.}
Section~\ref{sec:dirichlet} explains the obstruction to formalizing the
classical Feynman argument directly, and presents the proof through
\(\operatorname{sinc}^2\) together with its Lean implementation.
Section~\ref{sec:dirichlet_applications} derives the Dirichlet cutoff
and the corresponding Heaviside limit, as well as several quadratic
and product integral identities. Section~\ref{sec:lobachevsky}
establishes Lobachevsky's formula, beginning with cosine polynomials
and then passing to continuous periodic functions by uniform
approximation. The final section discusses possible extensions of the
development and concludes the paper.

\section{The Dirichlet Integral: Classical Proofs and Formalization}\label{sec:dirichlet}
The Dirichlet integral is one of the classical examples of an
improper integral and, in many undergraduate analysis courses, the
first example in which convergence is genuinely conditional. Its
apparently elementary integrand conceals the essential phenomenon:
the integral converges through oscillatory cancellation, although the
absolute value is not integrable.

\subsection{The classical Feynman argument and its formal obstruction}

A familiar evaluation of the Dirichlet integral begins by introducing
an exponential inverse factor. Following the method commonly
associated with Feynman \cite{schiff1999laplace}, define, for \(a>0\),
\[
    F(a)
    =
    \int_0^{+\infty}
        e^{-at}\operatorname{sinc}(t)\,dt.
\]
Differentiation under the integral sign is justified for every
\(a>0\), and yields
\[
    F'(a)
    =
    -\int_0^{+\infty}e^{-at}\sin t\,dt
    =
    -\frac{1}{1+a^2}.
\]
Since \(F(a)\to0\) as \(a\to+\infty\), it follows that
\[
    F(a)
    =
    \int_a^{+\infty}\frac{du}{1+u^2}
    =
    \frac{\pi}{2}-\arctan(a).
\]
In many textbook presentations, the proof is essentially concluded at
this point by letting \(a\) tend to zero under the integral sign:

\begin{align*}
    \lim_{a\to0^+}
    \int_0^{+\infty}
        e^{-at}\operatorname{sinc}(t)\,dt
    &\stackrel{?}{=}
    \int_0^{+\infty}
        \lim_{a\to0^+}
        e^{-at}\operatorname{sinc}(t)\,dt \\
    &=
    \int_0^{+\infty}\operatorname{sinc}(t)\,dt.
\end{align*}

The conclusion is correct, but it requires an additional estimate, uniform in the
exponential parameter. For \(a,T>0\), an integration by
parts gives the uniform estimate
\begin{equation*}
    |\int_T^{+\infty}
        e^{-at}\operatorname{sinc}(t)\,dt|
    \leq
    \frac{3}{T}.
\end{equation*}
which, together with dominated convergence on each fixed interval
\([0,T]\), justifies the passage \(a\to0^+\).

Thus, the argument is valid, but its usual presentation hides
an additional uniform tail estimate. Formalizing it would require
coordinating the limits \(a\to0^+\) and \(T\to+\infty\), together with
an integration-by-parts argument at an infinite endpoint.

We instead pass through \(\operatorname{sinc}^2\), which is absolutely
integrable on the positive half-line. The parameter can then
be removed directly by dominated convergence, and an identity between
truncated integrals of \(\operatorname{sinc}\) and
\(\operatorname{sinc}^2\) transfers the resulting value back to the
Dirichlet integral.

\subsection{An integrable detour through the squared sinc function}

Our proof first evaluates the absolutely convergent integral of
\(\operatorname{sinc}^2\), and then transfers its value to the
conditionally convergent Dirichlet integral. The squared sinc function
is integrable on the positive half-line: it is bounded near the origin
and satisfies
\[
    \operatorname{sinc}^2(t)\leq \frac{1}{t^2}
    \qquad (t\geq1).
\]

The connection between the two integrals follows from a change of
variables and an integration by parts. For every \(T>0\), one obtains
\begin{equation}\label{eq:sinc-sinc-square}
    \int_0^T\operatorname{sinc}(t)\,dt
    =
    \int_0^{T/2}\operatorname{sinc}^2(t)\,dt
    +
    \frac{T}{2}
        \operatorname{sinc}^2\left(\frac{T}{2}\right).
\end{equation}
The boundary term tends to zero, since
\[
    0\leq
    \frac{T}{2}
        \operatorname{sinc}^2\left(\frac{T}{2}\right)
    \leq \frac{2}{T}.
\]
It therefore remains to evaluate the integral of
\(\operatorname{sinc}^2\).

For \(x>0\), introduce the exponentially  integral
\[
    I(x)
    =
    \int_0^{+\infty}
        e^{-xt}\operatorname{sinc}^2(t)\,dt.
\]
Differentiating twice under the integral sign gives
\[
    I''(x)
    =
    \int_0^{+\infty}e^{-xt}\sin^2(t)\,dt
    =
    \frac{1}{2x}-\frac{x}{2(4+x^2)}.
\]
Together with the behavior of \(I\) and \(I'\) at infinity, this
determines
\begin{equation}\label{eq:damped-sinc-square}
    I(x)
    =
    \frac{x}{4}
        \log\left(\frac{x^2}{4+x^2}\right)
    +
    \arctan\left(\frac{2}{x}\right).
\end{equation}

Since \(\operatorname{sinc}^2\) is integrable, dominated convergence
allows the factor to be removed as \(x\to0^+\). Taking the
limit in \eqref{eq:damped-sinc-square} yields
\[
    \int_0^{+\infty}\operatorname{sinc}^2(t)\,dt
    =
    \frac{\pi}{2}.
\]
Finally, letting \(T\to+\infty\) in
\eqref{eq:sinc-sinc-square} gives
\[
    \lim_{T\to+\infty}
        \int_0^T\operatorname{sinc}(t)\,dt
    =
    \frac{\pi}{2}.
\]

The advantage of this route is that dominated convergence is applied
only to an absolutely integrable function. Conditional convergence is
handled separately through the identity
\eqref{eq:sinc-sinc-square}.\subsection{The Lean formalization}
The formalization keeps the two notions of integration separate.
The squared sinc function is treated as a Lebesgue-integrable function
on the positive half-line, whereas the Dirichlet integral is stated as
the convergence of bounded interval integrals.

The first step is the integrability of the squared sinc function. The
proof splits the positive half-line at \(1\), using continuity on the
bounded part and comparison with \(t^{-2}\) at infinity.

\begin{lstlisting}[language=lean]
lemma integrable_sinc_sq :
    IntegrableOn (fun (t : ℝ) ↦ (sinc t)^2) (Ioi 0)
\end{lstlisting}

The identity \eqref{eq:sinc-sinc-square} requires slightly more care.
Since integration by parts involves \(t\mapsto-1/t\), it is first
proved away from the origin. The lower endpoint is then sent to zero.

\begin{lstlisting}[language=lean]
lemma integral_sinc_sq_eq_dirichlet_bounded {a T : ℝ} (ha : 0 < a) (hT : a ≤ T) :
    (∫ t in a..T, Real.sinc t) = (∫ t in a / 2..T / 2, (Real.sinc t)^2)
                - (Real.sinc (a / 2))^2 * (a / 2) + (Real.sinc (T / 2))^2 * (T / 2)

lemma integral_sinc_zero_T (T : ℝ) (hT : T > 0) :
    (∫ t in 0..T, Real.sinc t) =
      (∫ t in 0..T / 2, (Real.sinc t)^2)
      + (Real.sinc (T / 2))^2 * (T / 2)
\end{lstlisting}

To evaluate the squared sinc integral, the development introduces the
integral \(I\) and its first two derivatives.

\begin{lstlisting}[language=lean]
def sinc_sq_times_exp (t : ℝ) : ℝ → ℝ :=
  fun x ↦ Real.exp (-x * t) * (Real.sinc t)^2

def neg_sinc_sq_times_id_exp (t : ℝ) : ℝ → ℝ :=
  fun x ↦ -(Real.sinc t)^2 * t * Real.exp (-x * t)

def sin_sq_times_exp (t : ℝ) : ℝ → ℝ :=
  fun x ↦ (Real.sin t)^2 * Real.exp (-x * t)

def integral_sinc_sq_times_exp (x : ℝ) : ℝ :=
  ∫ t in Ioi 0, sinc_sq_times_exp t x

def integral_neg_sinc_sq_times_id_exp (x : ℝ) : ℝ :=
  ∫ t in Ioi 0, neg_sinc_sq_times_id_exp t x

def integral_sin_sq_times_exp (x : ℝ) : ℝ :=
  ∫ t in Ioi 0, sin_sq_times_exp t x
\end{lstlisting}

For a fixed \(x>0\), differentiation under the integral sign is carried
out in a neighborhood where the parameter remains bounded below by
\(x/2\). This provides an integrable exponential majorant and yields
the two derivative identities.

\begin{lstlisting}[language=lean]
theorem hasDeriv_integral_sinc_sq_times_exp (x : ℝ) (hx : 0 < x) :
    HasDerivAt (integral_sinc_sq_times_exp)(integral_neg_sinc_sq_times_id_exp x) x

theorem hasDeriv_integral_neg_sinc_sq_times_id_exp (x : ℝ) (hx : 0 < x) :
    HasDerivAt (integral_neg_sinc_sq_times_id_exp) (integral_sin_sq_times_exp x) x
\end{lstlisting}

After evaluating the second derivative, the closed form is identified
from its derivative and its limit at infinity.

\begin{lstlisting}[language=lean]
theorem integral_sinc_sq_times_exp_eq (x : ℝ) (hx : 0 < x) :
    integral_sinc_sq_times_exp x = x / 4 * Real.log (x^2 / (4 + x^2)) + Real.arctan (2 / x)
\end{lstlisting}

Dominated convergence then removes the exponential factor. The resulting
evaluation is combined with \texttt{integral\_sinc\_zero\_T} to obtain
the Dirichlet integral.

\begin{lstlisting}[language=lean]
theorem integral_sinc_sq_eq_pi_div_two :
    ∫ t in Ioi 0, (Real.sinc t)^2 = π / 2


theorem integral_dirichlet :
    Tendsto (fun T ↦ ∫ t in 0..T, sinc t) atTop (�� (π / 2))
\end{lstlisting}

The final theorem therefore has the intended meaning: the Dirichlet
integral is a limit of truncated integrals, not the Lebesgue integral
of the non-integrable sinc function. The separate evaluation of
\(\operatorname{sinc}^2\) will also be used in the applications below.

\section{Further Integral Identities and the Dirichlet
Cutoff}\label{sec:dirichlet_applications}

The Dirichlet integral becomes particularly useful after scaling its
upper endpoint. We first use this observation to construct a smooth
approximation of the Heaviside function and then derive a cutoff
formula for integrable functions. Further trigonometric identities
will be considered in the following subsections.

\subsection{The Dirichlet cutoff and the Heaviside function}

Define the normalized primitive
\[
    D(x)
    =
    \frac12+\frac1\pi\int_0^x\operatorname{sinc}(t)\,dt
\]
and the Heaviside function
\[
    H(x)
    =
    \begin{cases}
        1,   & x>0,\\
        1/2, & x=0,\\
        0,   & x<0.
    \end{cases}
\]
The Dirichlet integral and the evenness of
\(\operatorname{sinc}\) imply that, for every \(x\in\mathbb R\),
\begin{equation}\label{eq:dirichlet-heaviside}
    \lim_{R\to+\infty}D(Rx)=H(x).
\end{equation}

The corresponding definitions and convergence theorem in Lean are:

\begin{lstlisting}[language=lean]
noncomputable def DirichletSin : ℝ → ℝ :=
  fun x↦1/2 + 1/π * ∫ t in  (0).. (x), sinc t


noncomputable def HeavisidePerso (x : ℝ) : ℝ :=
  if x > 0 then 1 else if x = 0 then 1/2 else 0


theorem lim_S_Rx (x : ℝ) : 
    Tendsto (fun R ↦ DirichletSin (R * x)) atTop (�� (HeavisidePerso x)) 
\end{lstlisting}

The formalization also establishes the standard analytic properties of
\(D\), including its continuity and boundedness on \(\mathbb R\).

\begin{lstlisting}[language=lean]
lemma DirichletSin_continuous :
    Continuous fun u ↦ DirichletSin (u)

theorem DirichletSinBounded :
    ∃ M, ∀ y, |DirichletSin y| ≤ M
\end{lstlisting}

This bound allows the pointwise convergence
\eqref{eq:dirichlet-heaviside} to be used under an integral. If
\(f\) is integrable and \(t\in\mathbb R\), then
\begin{equation}\label{eq:dirichlet-cutoff}
    \lim_{R\to+\infty}
        \int_{\mathbb R}
            f(a)D\bigl(R(a-t)\bigr)\,da
    =
    \int_t^{+\infty}f(a)\,da.
\end{equation}

\begin{lstlisting}[language=lean]
theorem Tendsto_Integral_DirichletSin_times_integrableFunction
    (f : ℝ → ℝ) (t : ℝ) (hf : Integrable (fun t ↦ f t)) :
    Tendsto (fun T : ℝ ↦ ∫ a, f a * DirichletSin (T * (a - t))) atTop (�� (∫ a in Ioi t, f a))
\end{lstlisting}

The development also proves the complex-valued analogue
\texttt{Tendsto\_\allowbreak Integral\_\allowbreak DirichletSin\_\allowbreak times\_\allowbreak integrableFunction'}
and a version restricted to the positive half-line,
\texttt{Tendsto\_\allowbreak Integral\_\allowbreak DirichletSin\_\allowbreak times\_\allowbreak integrableFunction\_\allowbreak zero'}.

\begin{lstlisting}[language=lean]
theorem Tendsto_Integral_DirichletSin_times_integrableFunction' (f : ℝ → ℂ) (t : ℝ) (hf : Integrable (fun t ↦ f t)) :
    Tendsto (fun T : ℝ ↦ ∫ a, f a * ↑(DirichletSin (T * (a - t)))) atTop (�� (∫ a in Ioi t, f a))
\end{lstlisting}

\begin{lstlisting}[language=lean]
theorem Tendsto_Integral_DirichletSin_times_integrableFunction_zero' (f : ℝ → ℂ) (t : ℝ) (hf : Integrable (fun t ↦ f t)) :
    Tendsto (fun T : ℝ ↦ ∫ a in Ioi 0, f a * ↑(DirichletSin (T * (a - t))))
             atTop (�� (∫ a in Ioi (max 0 t), f a))
\end{lstlisting}

These cutoff formulas will be used in the integral identities below.
\subsection{Quadratic trigonometric integrals}

The evaluation of the squared sinc integral immediately yields several
parameterized identities.

\subsubsection{Scaling the squared sinc integral}

For \(a>0\), the change of variables \(u=at\) gives
\begin{equation}\label{eq:scaled-sinc-square}
    \int_0^{+\infty}\operatorname{sinc}^2(at)\,dt
    =
    \frac{\pi}{2a}.
\end{equation}

\begin{lstlisting}[language=lean]
lemma integral_sinc_sq_scaled_of_pos (a : ℝ) (ha : 0 < a) :
    (∫ t in Ioi 0, (Real.sinc (a * t)) ^ 2) = (1 / a) * (Real.pi / 2)
\end{lstlisting}

\subsubsection{The quadratic sine kernel}

Since
\[
    \left(\frac{\sin(at)}{t}\right)^2
    =
    a^2\operatorname{sinc}^2(at)
\]
for \(t>0\), Equation~\eqref{eq:scaled-sinc-square}, together with
the symmetry in \(a\), gives
\begin{equation}\label{eq:quadratic-sine}
    \int_0^{+\infty}
        \left(\frac{\sin(at)}{t}\right)^2\,dt
    =
    \frac{\pi|a|}{2}.
\end{equation}

\begin{lstlisting}[language=lean]
theorem integral_sin_sq_div_sq (a : ℝ) :
    (∫ t in Ioi 0, (Real.sin (a * t) / t) ^ 2) = Real.pi * |a| / 2
\end{lstlisting}

The corresponding integrability statement is formalized as
\texttt{integrableOn\_\allowbreak sin\_\allowbreak sq\_\allowbreak div\_\allowbreak sq}.

\subsubsection{The cosine-difference kernel}

Using
\[
    1-\cos u=2\sin^2\left(\frac{u}{2}\right),
\]
Equation~\eqref{eq:quadratic-sine} yields
\begin{equation}\label{eq:cosine-difference}
    \int_0^{+\infty}
        \frac{1-\cos(at)}{t^2}\,dt
    =
    \frac{\pi|a|}{2}.
\end{equation}

\begin{lstlisting}[language=lean]
theorem integral_one_sub_cos_div_sq (a : ℝ) :
    (∫ t in Ioi 0, (1 - Real.cos (a * t)) / t ^ 2) = Real.pi * |a| / 2
\end{lstlisting}

Its integrability is recorded in
\texttt{integrableOn\_\allowbreak one\_\allowbreak sub\_\allowbreak cos\_\allowbreak div\_\allowbreak sq}. This last form is
particularly useful for evaluating products of sine functions.
\subsection{Products of sine and sinc functions}

The cosine-difference identity also gives closed forms for products of
sine and sinc functions.

\subsubsection{Products of sine functions}

For \(a,b\geq0\), the product-to-sum identity gives
\[
    2\sin(at)\sin(bt)
    =
    \bigl(1-\cos((a+b)t)\bigr)
    -
    \bigl(1-\cos((a-b)t)\bigr).
\]
Applying Equation~\eqref{eq:cosine-difference} yields
\[
\begin{aligned}
    \int_0^{+\infty}
        \frac{\sin(at)\sin(bt)}{t^2}\,dt
    &=
    \frac{\pi}{4}\bigl(a+b-|a-b|\bigr) \\
    &=
    \frac{\pi}{2}\min\{a,b\}.
\end{aligned}
\]

\begin{lstlisting}[language=lean]
theorem integral_sin_mul_sin_div_sq (a b : ℝ) (ha : 0 ≤ a) (hb : 0 ≤ b) :
    (∫ t in Ioi 0, Real.sin (a * t) * Real.sin (b * t) / t ^ 2) = Real.pi * min a b / 2
\end{lstlisting}

\subsubsection{Products of scaled sinc functions}

For \(a,b>0\),
\[
    \operatorname{sinc}(at)\operatorname{sinc}(bt)
    =
    \frac{1}{ab}
    \frac{\sin(at)\sin(bt)}{t^2}
\]
almost everywhere on the positive half-line. Consequently,
\[
\begin{aligned}
    \int_0^{+\infty}
        \operatorname{sinc}(at)\operatorname{sinc}(bt)\,dt
    &=
    \frac{\pi\min\{a,b\}}{2ab} \\
    &=
    \frac{\pi}{2\max\{a,b\}}.
\end{aligned}
\]

\begin{lstlisting}[language=lean]
theorem integral_sinc_mul_sinc (a b : ℝ) (ha : 0 < a) (hb : 0 < b) :
    (∫ t in Ioi 0, Real.sinc (a * t) * Real.sinc (b * t)) = Real.pi / (2 * max a b)
\end{lstlisting}

These identities provide the trigonometric integral evaluations used
in the proof of Lobachevsky's formula.

\section{Lobachevsky's Integral Formula}\label{sec:lobachevsky}

The identities established in the preceding section are instances of
a broader phenomenon: when the squared sinc function is multiplied by
a suitably symmetric periodic function, the integral over the entire
positive half-line reduces to an integral over a single half-period.
Lobachevsky's formula makes this principle precise.

\subsection{Historical background and statement}

Although Nikolai Lobachevsky is primarily remembered for
non-Euclidean geometry, he also worked on probability, trigonometric
series, and definite integrals. TThe formulas now bearing his name originate in his 1852 paper on methods for evaluating improper integrals \cite[p. 277]{lobachevsky1852}

Later generalizations have considered higher powers of the sinc
function \cite{jolany2018} and interpretations based on Fourier
analysis, Parseval-type identities, and the Shannon sampling basis
\cite{cai2022}. We formalize the classical formula for the absolutely
integrable squared sinc kernel.

\begin{theorem}[Lobachevsky's integral formula]
\label{thm:lobachevsky}
Let \(f\colon\mathbb R\to\mathbb R\) be continuous and satisfy
\[
    f(x+\pi)=f(x)
    \qquad\text{and}\qquad
    f(\pi-x)=f(x)
\]
for every \(x\in\mathbb R\). Then
\[
    \int_0^{+\infty}
        \operatorname{sinc}^2(x)f(x)\,dx
    =
    \int_0^{\pi/2}f(x)\,dx.
\]
\end{theorem}

The symmetry assumptions imply that \(f\) is even and suggest
approximation by cosine polynomials
\[
    p(x)=\sum_{n=0}^{N}a_n\cos(2nx).
\]
The proof first establishes the formula for each cosine mode and then
passes to a uniform limit.

\subsection{The proof for cosine polynomials}

For \(n\geq1\), the identity
\[
\begin{aligned}
    \operatorname{sinc}^2(x)\cos(2nx)
    =
    \frac14\bigg(
      &\frac{1-\cos(2(n+1)x)}{x^2}
       +\frac{1-\cos(2(n-1)x)}{x^2}\\
      &-2\frac{1-\cos(2nx)}{x^2}
    \bigg)
\end{aligned}
\]
and Equation~\eqref{eq:cosine-difference} give
\[
    \int_0^{+\infty}
        \operatorname{sinc}^2(x)\cos(2nx)\,dx
    =0.
\]
For \(n=0\), the integral is \(\pi/2\). These are also the values of
the corresponding integrals over \([0,\pi/2]\):
\[
    \int_0^{\pi/2}\cos(2nx)\,dx
    =
    \begin{cases}
        \pi/2,& n=0,\\
        0,& n\geq1.
    \end{cases}
\]

The nonconstant mode calculation is formalized as follows.

\begin{lstlisting}[language=lean]
lemma integral_sinc_sq_mul_cos_two_nat (n : ℕ) (hn : 0 < n) :
    (∫ x in Set.Ioi 0, (Real.sinc x) ^ 2 * Real.cos (2 * (n : ℝ) * x)) = 0
\end{lstlisting}

For
\[
    p(x)=\sum_{n=0}^{N}a_n\cos(2nx),
\]
linearity now gives
\begin{equation}\label{eq:lobachevsky-cosine-polynomial}
    \int_0^{+\infty}
        \operatorname{sinc}^2(x)p(x)\,dx
    =
    \int_0^{\pi/2}p(x)\,dx.
\end{equation}

\begin{lstlisting}[language=lean]
def cosinePolynomial (N : ℕ) (a : ℕ → ℝ) (x : ℝ) : ℝ :=
  ∑ n ∈ Finset.range (N + 1), a n * Real.cos (2 * (n : ℝ) * x)

lemma lobachevsky_cosinePolynomial (N : ℕ) (a : ℕ → ℝ) :
    (∫ x in Set.Ioi 0, (Real.sinc x) ^ 2 * cosinePolynomial N a x) =
        ∫ x in (0 : ℝ)..Real.pi / 2, cosinePolynomial N a x
\end{lstlisting}

\subsection{Uniform approximation on the additive circle}

It remains to approximate every continuous function satisfying the
hypotheses of Theorem~\ref{thm:lobachevsky} by cosine polynomials.
More precisely, for every \(\varepsilon>0\), we seek
\[
    p(x)=\sum_{n=0}^{N}a_n\cos(2nx)
\]
such that
\[
    |f(x)-p(x)|<\varepsilon
    \qquad\text{for all }x\in\mathbb R.
\]

Periodicity allows \(f\) to descend to a continuous function on
\[
    \operatorname{AddCircle}\pi=\mathbb R/\pi\mathbb Z.
\]
Mathlib's theorem
\texttt{span\_\allowbreak fourier\_\allowbreak closure\_\allowbreak eq\_\allowbreak top}, from
\texttt{Mathlib.\allowbreak Analysis.\allowbreak Fourier.\allowbreak AddCircle}, states that finite
Fourier sums are dense among continuous complex-valued functions on
this circle. In the formal proof, we denote this local density statement by
\texttt{hS\_dense}.

The reflection condition, together with periodicity, implies that
\(f\) is even. Taking the real even part of a Fourier approximation
and pairing the frequencies \(k\) and \(-k\) removes the sine terms
and produces the required cosine polynomial.

\begin{lstlisting}[language=lean]
lemma exists_cosinePolynomial_uniform_approx_of_reflection {f : ℝ → ℝ} (hf_cont : Continuous f) 
        (hf_periodic : Function.Periodic f Real.pi)
        (hf_reflection : ∀ x : ℝ, f (Real.pi - x) = f x) {ε : ℝ} (hε : 0 < ε) :
    ∃ N : ℕ, ∃ a : ℕ → ℝ, ∀ x : ℝ, |f x - cosinePolynomial N a x| < ε
\end{lstlisting}

\subsection{The uniform limit and the formal theorem}

Define
\[
    L(g)=\int_0^{+\infty}\operatorname{sinc}^2(x)g(x)\,dx,
    \qquad
    R(g)=\int_0^{\pi/2}g(x)\,dx.
\]
We conclude by contradiction. Suppose that \(L(f)\neq R(f)\), and set
\[
    \eta=\frac{|L(f)-R(f)|}{2\pi}>0.
\]
Choose a cosine polynomial \(p\) satisfying
\[
    |f(x)-p(x)|<\eta
    \qquad\text{for every }x\in\mathbb R.
\]
Uniform approximation gives
\[
    |L(f)-L(p)|\leq\frac{\pi}{2}\eta,
    \qquad
    |R(p)-R(f)|\leq\frac{\pi}{2}\eta.
\]
Since Equation~\eqref{eq:lobachevsky-cosine-polynomial} gives
\(L(p)=R(p)\), we obtain
\[
\begin{aligned}
    |L(f)-R(f)|
    &\leq |L(f)-L(p)|+|R(p)-R(f)|\\
    &\leq \pi\eta
     =\frac12|L(f)-R(f)|,
\end{aligned}
\]
a contradiction. Thus \(L(f)=R(f)\).

In Lean, the two estimates are supplied by
\texttt{abs\_integral\_sinc\_sq\_mul\_sub\_le} and
\texttt{abs\_intervalIntegral\_sub\_le}. Together with the cosine
approximation theorem, they yield the final result.

\begin{lstlisting}[language=lean]
theorem lobachevsky_integral_formula {f : ℝ → ℝ} (hf_cont : Continuous f)
        (hf_periodic : Function.Periodic f Real.pi)
        (hf_reflection : ∀ x : ℝ, f (Real.pi - x) = f x) :
    (∫ x in Set.Ioi 0, (Real.sinc x) ^ 2 * f x) = ∫ x in (0 : ℝ)..Real.pi / 2, f x
\end{lstlisting}

\section{New Directions and Conclusion}
In this paper, we have formalized the Dirichlet integral as the limit
of its truncated integrals, without identifying it with a Lebesgue
integral of the non-integrable sinc function. The detour through
\(\operatorname{sinc}^2\) provides both a proof of the Dirichlet
integral and an independent evaluation of an absolutely convergent
integral. From these two results, we derived the Dirichlet cutoff and
several quadratic and bilinear trigonometric identities. We then used
the same analytic foundation, together with Fourier approximation on
\(\operatorname{AddCircle}\pi\), to obtain Lobachevsky's integral
formula.

The present development suggests several natural extensions. At the
level of integration theory, the Dirichlet integral illustrates the
need for a convenient formal treatment of conditionally convergent
improper integrals. In this paper, such an integral is expressed
directly as a limit of integrals over bounded intervals. A more
systematic interface for improper integrals could collect the usual
operations (changes of variables, integration by parts, and
comparison of different truncations) and make similar formalizations more direct.

The results on the Dirichlet cutoff also point toward further
applications in Fourier analysis. Its convergence to the Heaviside
function is a basic example of the way oscillatory kernels recover
discontinuous functions. A natural continuation would be to use these
results in formal proofs of Fourier inversion and of convergence
theorems at points of discontinuity.

The most immediate extension of the Lobachevsky part concerns higher
even powers of the sinc function. Jolany developed formulas for
integrals involving \(\operatorname{sinc}^{2n}\) and symmetric
periodic weights \cite{jolany2018}. Already for the fourth power, the
result takes the form
\[
    \int_0^{+\infty}\operatorname{sinc}^4(x)f(x)\,dx
    =
    \int_0^{\pi/2}f(x)\,dx
    -
    \frac{2}{3}
    \int_0^{\pi/2}\sin^2(x)f(x)\,dx.
\]
The difference with the squared sinc formula is significant. For
\(\operatorname{sinc}^2\), only the constant cosine mode contributes
to the integral, whereas higher powers retain several low-frequency
modes. Formalizing these extensions would therefore require a broader
collection of mode identities, but the approximation on
\(\operatorname{AddCircle}\pi\) developed here should continue to
provide the passage from cosine polynomials to continuous periodic
functions.

Finally, the Dirichlet integral provides an elegant real-analytic
route to Laplace inversion that avoids the residue theorem, and more
generally supports the development of the theory of Laplace
transforms. This approach is used in the companion formalization
\cite{GoldbergVinciguerraLaplace}.

\section{Acknowledgements}
 
The authors would like to thank Professor Yuval Filmus for being at the origin of this
project and for encouraging its development. We are also grateful to Ashvni Narayanan for through and insightful discussions.
The authors would also like to thank Professor Keith Conrad for highlighting the original Lobachevski's paper.

\end{document}